\documentstyle[prl,preprint,aps,epsf]{revtex}
\begin{document}

\title{Absorption Cross Section for S-wave massive Scalar}
\author{Eylee Jung\footnote{Email:eylee@mail.kyungnam.ac.kr}, SungHoon Kim\footnote{Email:shoon@mail.kyungnam.ac.kr}, and
D. K. Park\footnote{Email:dkpark@hep.kyungnam.ac.kr 
}, }
\address{Department of Physics, Kyungnam University, Masan, 631-701, Korea}

\maketitle

\maketitle
\begin{abstract}
We examine the absorption cross section of the massive scalar field
for the higher-dimensional extended object. 
Adopting the usual quantum mechanical matching conditions between
the asymptotic and near-horizon solutions in radial equation, 
we check whether or not the 
universal property of the absorption cross section, which is that the 
low-energy cross section is proportional to the surface area of horizon, is 
maintained when the mass effect is involved. 
It is found that the mass effect in general does not break the 
universal property of the cross section if particular conditions are 
required to the spacetime geometry. However, the mass-dependence of 
the cross section is very sensitive to the spacetime property in the 
near-horizon regime.
\end{abstract}

\newpage
It is well-known that the low-energy absorption cross section of the 
massless particle for the black hole is proportional to the surface area of the 
horizon\cite{sta74,gibb75,page76,emp98,park00,das97}. 
This universal property is shown to be 
maintained for the higher-dimensional objects such as extremal strings and 
black $p$-brane\cite{emp98}. 
Furthermore, adopting the usual
quantum mechanical matching conditions the authors in Ref.\cite{park00} have 
argued that the genuine physical reason for the occurrence of the universal
property is the independence of the matching point between the asymptotic
and near-horizon solutions. 

In this letter we will examine whether the universal property of the low-energy
absorption cross section is maintained or not in the higher-dimensional 
extended system when the mass effect of the scalar field is involved. In order
to proceed we will adopt the quantum mechanical matching conditions
\begin{eqnarray}
\label{match2}
\phi_{\omega}^{\infty}(R)&=&\phi_{\omega}^0(R)   \\   \nonumber
\frac{d}{d R} \phi_{\omega}^{\infty}(R)&=& \frac{d}{d R}
\phi_{\omega}^0(R)
\end{eqnarray}
which is introduced in Ref.\cite{park00}. In Eq.(\ref{match2}) 
$\phi_{\omega}^{\infty}$ and $\phi_{\omega}^0$ are the asymptotic and 
near-horizon solutions respectively of the massive scalar field and 
thus, Eq.(\ref{match2}) is the matching between them at arbitrary location
$r = R$. 

We will follow Ref.\cite{park00} to check the universality 
when the scalar field is massive. The mass effect in the scalar field 
requires an explicit $r$-dependence of the time-time component of the
metric for the derivation of $\phi_{\omega}^0$. (See Eq.(\ref{assump})) 
The mass-dependence of the
absorption cross section is very sensitive to the $r$-dependence of the 
metric. In spite of it the mass of the scalar particle does not break
the universal property of the absorption cross section if particular 
conditions are required.

The spacetime generated by the higher-dimensional object is assumed to 
be 
\begin{equation}
\label{metric}
ds^2 = \gamma_{\mu \nu} dx^{\mu} dx^{\nu} + f(r) dr^2 + r^2 h(r) 
d\Omega_{n+1}
\end{equation}
through this letter,
where $\mu, \nu = 0, 1, \cdots p$. 
Let us consider the minimally coupled massive scalar in this background, 
which should satisfy
\begin{equation}
\label{minimal}
(\Delta_A \Delta^A - m^2)  \Phi = 0.
\end{equation}
If we assume $\Phi = e^{-i \omega t} \phi_{\omega} (r)$ which is valid
for the low-energy $s$-wave, and introduce a ``tortose'' coordinate 
$r_{\ast}$ as following, 
\begin{equation}
\label{tortose}
dr_{\ast} = dr \sqrt{- \gamma^{tt} f(r)},
\end{equation}
one can show straightforwardly that Eq.(\ref{minimal}) reduces to the 
following Schr\"{o}dinger-like equation
\begin{equation}
\label{schro}
\left[
      -\frac{d^2}{d r_{\ast}^2} + V(r) \right] \psi = \omega^2 \psi
\end{equation}
where $\gamma(r) \equiv det \gamma_{\mu \nu}$ and
\begin{eqnarray}
\label{schro1}
\psi(r)&=&\sqrt{U} \phi_{\omega}  \hspace{2.0cm} 
U(r)=\sqrt{\gamma \gamma^{tt} \left\{r^2 h(r)\right\}^{n+1}}  \\  \nonumber
V(r)&=&V_0(r) - \frac{m^2}{\gamma^{tt}} \hspace{1.0cm}
V_0(r)=\frac{1}{\sqrt{U}}
         \frac{d^2\sqrt{U}}{(d r_{\ast})^2}.
\end{eqnarray}

Usually the ``tortose'' coordinate $r_{\ast}$ goes to $\pm\infty$ in the 
asymptotic and near-horizon regions of $r$ and the potential $V_0(r)$ makes 
a barrier which separates these two regions. If, for example, we consider the
usual $4d$ Schwarzschild spacetime by choosing 
$f(r) = - \gamma^{tt} = (1 - r_H / r)^{-1}$, the ``tortose'' coordinate 
$r_{\ast}$ and the potential $V_0(r)$ reduce to 
$r_{\ast} = r + r_H \ln (r - r_H)$ and $V_0(r) = (r_H / r^3) (1 - r_H / r)$.
Fig. 1 shows $r_{\ast}$-dependence of $V_0(r)$ in this simple example when
$r_H = 1$. From Fig. 1 we understand that $V_0(r)$ makes a barrier between 
the asymptotic and near-horizon regions.

We will solve Eq.(\ref{schro}) in the asymptotic region ($r \sim \infty$) and 
near-horizon region ($r \sim 0$) separately. Matching them using 
Eq.(\ref{match2}), we will derive the absorption cross section. Firstly, 
let us consider Eq.(\ref{schro}) in the asymptotic region with assumption that 
the geometry is asymptotic flat for simplicity;
\begin{eqnarray}
\label{asymp}
\lim_{r \rightarrow \infty} \gamma_{\mu \nu}&=&\eta_{\mu \nu}
                                                              \\   \nonumber
\lim_{r \rightarrow \infty} f(r)&=&\lim_{r \rightarrow \infty} h(r) = 1.
\end{eqnarray}


Using Eq.(\ref{asymp}), Eq.(\ref{schro}) reduces to 
\begin{equation}
\label{asradial1}
\frac{d^2 u_{\infty}}{d r^2} + \frac{1}{r} \frac{d u_{\infty}}{d r}
+ \left( \omega^2 - m^2 - \frac{n^2}{4 r^2} \right) u_{\infty} = 0
\end{equation}
in this region where $u_{\infty} \equiv \psi_{\infty} / \sqrt{r}$.
Here the subscript denotes the region we consider for the solution
of Eq.(\ref{schro}). 
Eq.(\ref{asradial1}) is easily solved in terms of Bessel function 
and hence we can derive
\begin{equation}
\label{solution2}
\phi_{\omega}^{\infty} \equiv \sqrt{\frac{r}{U}} u_{\infty} = 
\frac{1}{(\omega v r)^{\frac{n}{2}}} 
\left[A J_{\frac{n}{2}} (\omega v r) + B J_{-\frac{n}{2}} (\omega v r)\right]
\end{equation}
where $v = \sqrt{1 - m^2 / \omega^2}$. 

Now we define the flux of the massive scalar as 
\begin{equation}
\label{flux-def}
{\cal F} = \frac{1}{2 i}
           \frac{U}{\sqrt{- \gamma^{tt} f(r)}}
           \left( \phi_{\omega}^{\ast} \frac{d \phi_{\omega}}{d r} - 
                  \phi_{\omega} \frac{d \phi_{\omega}^{\ast}}{d r} \right).
\end{equation}
Inserting (\ref{solution2}) into Eq.(\ref{flux-def}) yields the incomong 
flux in the form;
\begin{equation}
\label{influx1}
{\cal F}_{\infty}^{in} = \frac{-1}{2 \pi (\omega v)^n}
\left[ |A|^2 + |B|^2 + A^{\ast} B e^{-i \frac{n}{2} \pi} + 
       A B^{\ast} e^{i \frac{n}{2} \pi} \right].
\end{equation}
When deriving Eq.(\ref{influx1}) we used only the incoming wave,
{\it i.e.} $e^{-i \omega v r}$, in the asymptotic formula of 
Bessel function.


Now we will solve Eq.(\ref{schro}) in the near-horizon region.
Following Ref.\cite{emp98,park00} we take a following assumption
\begin{eqnarray}
\label{assump}
\lim_{r \rightarrow 0} U&\sim& S r^{a - b}   \\   \nonumber
\lim_{r \rightarrow 0} \sqrt{- \gamma^{tt} f}&\sim& \frac{T}{r^{b + 1}}
                                                     \\   \nonumber
\lim_{r \rightarrow 0} \gamma^{tt} &\sim& - \frac{W}{r^{2 c}}
\end{eqnarray}
where $S$, $T$ and $W$ are some constant parameters. Especially, the parameter
$S$ is proportional to the area of the absorption hypersurface\footnote{In the 
black hole spacetime this is same with area of the horizon surface.}. Thus, the 
universality 
means that the absorption cross section 
for the low-energy massless particle is proportional to the parameter $S$. 
For example, the low-energy cross section for the massless scalar particle 
is found to be $\sigma_L = \Omega_{n+1} S$\cite{emp98,park00} when $a = b$, 
where $\Omega_{n+1}$ is an surface area of $S^{n+1}$, which exactly coincides
with the area of the absorption hypersurface.

Making use of Eq.(\ref{assump}) we can transform Eq.(\ref{schro}) into
\begin{equation}
\label{horadial1}
\frac{d^2 \chi}{d y^2} + \frac{1}{y} \frac{d \chi}{d y} + 
\left[ 1 + V_1(y) + V_2(y) \right] \chi = 0
\end{equation}
in the $r \sim 0$ region, where 
\begin{eqnarray}
\label{horadial2}
y&=& \frac{\omega T}{b r^b}
\hspace{3.5cm}
\chi(r) = r^{\frac{b}{2}} \psi_0    \\   \nonumber
V_1(y)&=&-\frac{m^2}{W \omega^2}
         \left(\frac{\omega^2 T^2}{b^2 y^2}\right)^{\frac{c}{b}}
\hspace{1.3cm}
V_2(y) = - \frac{a^2}{4 b^2} \frac{1}{y^2}.
\end{eqnarray}
It seems to be impossible to solve Eq.(\ref{horadial1}) analytically if 
both of $V_1(y)$ and $V_2(y)$ are present. Thus we should take an approximation
for the analytical approach.

If $0 < b < c$, $V_2(y)$ is much greater than $V_1(y)$, which makes 
$\chi$ to be proportional to $H_{\frac{a}{2 b}}^{(2)}(y)$ where 
$H_{\nu}^{(2)}$ is usual Hankel function. Thus we can take a solution as
\begin{equation}
\label{solution3}
\phi_{\omega}^0 = 
\frac{1}{(\omega r)^{\frac{a}{2}}} H_{\frac{a}{2 b}}^{(2)}
\left( \frac{\omega T}{b r^b} \right)
\end{equation}
in this region.

If $b = c$, $V_1(y)$ and $V_2(y)$ are almost same order, which results in
\begin{equation}
\label{solution4}
\phi_{\omega}^0 =
\frac{1}{(\omega r)^{\frac{a}{2}}} H_{\nu}^{(2)}
\left( \frac{\omega T}{b r^b} \right)
\end{equation}
where
\begin{equation}
\label{expl1}
\nu = \sqrt{\frac{a^2}{4 b^2} + \frac{m^2 T^2}{W b^2}}.
\end{equation}

If $0 < c < b$, $V_1(y)$ is a dominant term in the potential. In this case the 
solution of Eq.(\ref{horadial1}) cannot be solved in general. If $b = 2 c$,
however, we can solve Eq.(\ref{horadial1}), which results in
\begin{equation}
\label{solution5}
\phi_{\omega}^0 = \frac{1}{(\omega r)^{\frac{a - b}{2}}}
\left[F_{\ell} \left( \eta, \frac{\omega T}{b r^b} \right)
   + i G_{\ell} \left( \eta, \frac{\omega T}{b r^b} \right) \right]
\end{equation}
where $F_{\ell}(\eta, z)$ and $G_{\ell}(\eta, z)$ are the Coulomb 
wave functions and 
\begin{equation}
\label{revise1}
\eta = \frac{m^2 T}{2 W \omega b}, 
\hspace{2.0cm}
\ell = \frac{a - b}{2 b}.
\end{equation} 
In Eq.(\ref{solution5}) the coefficients of the Coulomb wave functions are 
chosen from a condition that we have a pure incoming wave at 
$r \sim 0$ region. In the following we will compute the low-energy absorption
cross section for $b < c$, $b = c$, and $b = 2c$ separately.


At $0 < b < c$, the solution in near-horizon region is 
Eq.(\ref{solution3}).
Then it is easy to show that the incident flux for $\phi_{\omega}^0$ is 
\begin{equation}
\label{influx2}
{\cal F} = \frac{1}{2 i} \frac{U}{\sqrt{-\gamma^{tt} f}}
\left( \phi_{\omega}^{0 \ast} \frac{d \phi_{\omega}^0}{d r}
      - \phi_{\omega}^0 \frac{d \phi_{\omega}^{0 \ast}}{d r} \right)
= \frac{2 b S}{\pi \omega^a T}.
\end{equation}
Thus the absorption cross section defined as 
\begin{equation}
\label{cross1}
\sigma \equiv \frac{(2 \pi)^{n+1}}{\omega^{n+1} \Omega_{n+1}}
\Bigg | \frac{{\cal F}_0^{in}}{{\cal F}_{\infty}^{in}} \Bigg |
\end{equation}
becomes 
\begin{equation}
\label{cross2}
\sigma = \frac{4 (2 \pi)^{n+1} b S v^n}{\omega^{a+1} T \Omega_{n+1}}
\frac{1}{|A|^2 + |B|^2 + A^{\ast} B e^{-i \frac{n}{2} \pi} + A B^{\ast}
e^{i \frac{n}{2} \pi}}
\end{equation}
where $\Omega_{n+1}$ is surface area of $S^{n+1}$, {\it i.e.}
$\Omega_{n+1} =  2 \pi^{1 + n/2} / \Gamma(1+n/2)$.

Now let us consider the matching between $\phi_{\omega}^{\infty}$ and 
$\phi_{\omega}^0$. In Ref.\cite{emp98} author uses 
\begin{equation}
\label{match1}
\lim_{r \rightarrow 0} \phi_{\omega}^{\infty} = 
\lim_{r \rightarrow \infty} \phi_{\omega}^0.
\end{equation}
This condition requires implicitly the assumption that there
exists an intermediate region where $\phi_{\omega}^0$ and 
$\phi_{\omega}^{\infty}$ can be matched. However, it is not clear at 
least for us to take this assumption {\it ab initio}.

Instead of this the authors in Ref.\cite{park00} took Eq.(\ref{match2}) as
matching conditions. Thus we do not need to assume the existence of the 
intermediate region
from the beginning.
If we solve Eq.(\ref{match2}) with the asymptotic solution (\ref{solution2})
and the near-horizon solution (\ref{solution3}), 
the coefficients $A$ and $B$ become
\begin{eqnarray}
\label{coeff1}
A&=& (-1)^{\frac{n+1}{2}}
    \frac{\pi (\omega R)^{\frac{n - a}{2}} v^{\frac{n}{2}}}{2}
    \Bigg[\frac{-n + a}{2} J_{-\frac{n}{2}}(\omega v R)
          H_{\frac{a}{2 b}}^{(2)} \left(\frac{\omega T}{b R^b}\right) 
                                                            \\  \nonumber
& & \hspace{2.0cm} + 
          \omega v R J_{-\frac{n}{2}}^{\prime}(\omega v R) 
          H_{\frac{a}{2 b}}^{(2)} \left(\frac{\omega T}{b R^b}\right) + 
          \frac{\omega T}{R^b} J_{-\frac{n}{2}}(\omega v R)
          H_{\frac{a}{2 b}}^{(2) \prime} \left(\frac{\omega T}{b R^b}\right)
                                                         \Bigg]
                                                               \\ \nonumber
B&=& (-1)^{\frac{n-1}{2}}
     \frac{\pi (\omega R)^{\frac{n - a}{2}} v^{\frac{n}{2}}}{2}
     \Bigg[\frac{-n + a}{2} J_{\frac{n}{2}}(\omega v R)
          H_{\frac{a}{2 b}}^{(2)} \left(\frac{\omega T}{b R^b}\right) 
                                                              \\  \nonumber
& & \hspace{2.0cm} + 
          \omega v R J_{\frac{n}{2}}^{\prime}(\omega v R) 
          H_{\frac{a}{2 b}}^{(2)} \left(\frac{\omega T}{b R^b}\right) + 
          \frac{\omega T}{R^b} J_{\frac{n}{2}}(\omega v R)
          H_{\frac{a}{2 b}}^{(2) \prime} \left(\frac{\omega T}{b R^b}\right)
                                                         \Bigg]  
\end{eqnarray}
where the prime denotes the differentiation with respect to the argument. 
Inserting Eq.(\ref{coeff1}) into Eq.(\ref{cross2}) one can compute the 
absorption cross section $\sigma^{(1)}$ straightforwardly. 

In order to show that the
the low-energy cross section is independent of the matching point, we plot
the $\omega$-dependence of $\sigma^{(1)}$ in Fig.2, which indicates that
in $\omega \sim m$ region $\sigma^{(1)}$ is independent of $R$.\footnote{
In order to apply our low-energy formulation we should require that the mass
of the scalar particle is not large.} Thus the 
low-energy universality seems to be maintained when $ 0 < b < c$ if 
$m$ is not too large. To show this 
more explicitly, we compute the coefficients $A$ and $B$ in the low energy
limit using the asymptotic formulae of Bessel and Hankel functions which 
results in
\begin{equation}
\label{coeff2}
A=\frac{i}{\pi} 2^{\frac{n}{2}} \Gamma \left( \frac{a}{2 b} \right)
    \Gamma \left( 1 + \frac{n}{2} \right)
    \left( \frac{2b}{T \omega^{b+1}} \right)^{\frac{a}{2 b}}
                                                        \hspace{1.0cm} 
B=0
\end{equation}
at the leading order. It is worthwhile noting that the $R$-dependence 
disappears in $A$ and $B$ 
as indicated before as a 
physical origin of the universality in
the low-energy cross-section. Computing the low-energy cross section by making 
use of  
Eq.(\ref{coeff2}), one can obtain easily
\begin{equation}
\label{cross3}
\sigma_L^{(1)} = \frac{\pi}{\Gamma^2 \left( \frac{a}{2 b} \right)}
\Omega_{n+1} S \left( \frac{\omega T}{2 b} \right)^{\frac{a}{b} -1} v^n
\end{equation}    
where $L$ in subscript stands for low-energy limit. If $a = b$, 
$\sigma_L^{(1)}$ becomes simply
\begin{equation}
\label{cross4}
\sigma_L^{(1)} = \Omega_{n+1} S v^n
\end{equation}
which indicates that the mass in scalar particle decreases the absorption 
cross section. A similar decreasing behavior of the absorption cross
section with respect to $m$ was also found by Unruh\cite{unruh76}.  

The authors in Ref.\cite{park00} applied
the matching condition (\ref{match2}) to the case of the fixed 
scalar\cite{kol97}, where the low-energy absorption cross section does not
obey the universality. The authors in Ref.\cite{kol97} computed the low-energy
cross section $\sigma_s$ by matching $\phi_{\omega}^0$ and 
$\phi_{\omega}^{\infty}$ through the solution in the intermediate region as 
Unruh did in his seminal paper\cite{unruh76} and obtained 
$\sigma_s = 2 \pi \omega^2$. If one uses, however, the matching condition
(\ref{match2}), one gets $\sigma_s = 2 \pi \omega^2 R^2 / (R-1)^2$. The 
explicit $R$-dependence indicates the non-universality, but its 
$\omega$-dependence is correct. This result may give us a confidence to 
use (\ref{match2}) to examine the $\omega$-dependence of the absorption 
cross section in the high-energy limit. 

If one takes, for example, $\omega \rightarrow \infty$ limit in the 
coefficients $A$ and $B$ of Eq.(\ref{coeff1}), it is easy to show that
these coefficients are explicitly dependent on the matching point $R$. 
Then, it is easy to show that the high-energy absorption cross section becomes
\begin{equation}
\label{hcross1}
\sigma_H^{(1)} = 
\frac{(2\pi / \omega)^{n+1}}{\Omega_{n+1} R^{n-a}}
\frac{4 S / T}{\left[ \sqrt{ \frac{v R^{b+1}}{T}} - \sqrt{\frac{T}{v R^{b+1}}}
                                             \right]^2}.
\end{equation}
The appearance of $R$ in Eq.(\ref{hcross1}) indicates that the high energy
cross section loses the universality property. However, the 
$\omega$-dependence of $\sigma_H^{(1)}$, {\it i.e.} 
$\sigma_H^{(1)} \propto \omega^{-(n+1)}$ exhibits a decreasing behavior.
A similar decreasing behavior was shown in Ref.\cite{cve00} by adopting a
numerical methods and in Ref.\cite{man00,park01} by analyzing a modified Mathieu
equation. 


Now let us consider the case of $b=c$. In this case 
the asymptotic and near-horizon solutions 
are (\ref{solution2}) and 
(\ref{solution4}) respectively. The difference of order of Hankel
function in (\ref{solution4}) from (\ref{solution3}) for $0 < b < c$ 
does not change the incoming flux of the near-horizon region because the 
order of Hankel function is only involved as a phase factor in the
asymptotic formula. Thus, the low-energy absorption cross section has a 
same expression with Eq.(\ref{cross2}). The difference of the low-energy
cross section, however, from that for $0 < b < c$ case arises due to
the matching between the asymptotic and near-horizon solutions. Applying
the matching condition (\ref{match2}) we obtain
\begin{eqnarray}
\label{coeff4}   
A&=& (-1)^{\frac{n+1}{2}}
    \frac{\pi (\omega R)^{\frac{n - a}{2}} v^{\frac{n}{2}}}{2}
    \Bigg[\frac{-n + a}{2} J_{-\frac{n}{2}}(\omega v R)
          H_{\nu}^{(2)} \left(\frac{\omega T}{b R^b}\right)
                                                            \\  \nonumber
& & \hspace{2.0cm} +
          \omega v R J_{-\frac{n}{2}}^{\prime}(\omega v R)
          H_{\nu}^{(2)} \left(\frac{\omega T}{b R^b}\right) +
          \frac{\omega T}{R^b} J_{-\frac{n}{2}}(\omega v R)
          H_{\nu}^{(2) \prime} \left(\frac{\omega T}{b R^b}\right)
                                                         \Bigg]
                                                               \\ \nonumber
B&=& (-1)^{\frac{n-1}{2}}
     \frac{\pi (\omega R)^{\frac{n - a}{2}} v^{\frac{n}{2}}}{2}
     \Bigg[\frac{-n + a}{2} J_{\frac{n}{2}}(\omega v R)
          H_{\nu}^{(2)} \left(\frac{\omega T}{b R^b}\right)
                                                              \\  \nonumber
& & \hspace{2.0cm} +
          \omega v R J_{\frac{n}{2}}^{\prime}(\omega v R)
          H_{\nu}^{(2)} \left(\frac{\omega T}{b R^b}\right) +
          \frac{\omega T}{R^b} J_{\frac{n}{2}}(\omega v R)
          H_{\nu}^{(2) \prime} \left(\frac{\omega T}{b R^b}\right)
                                                         \Bigg].
\end{eqnarray}
If we take $\omega \rightarrow m \sim 0$ limit
in Eq.(\ref{coeff4}), it is 
easy to show that the coefficients $A$ and $B$ are explicitly dependent
on $R$ unlike massless and $0 < b < c$ cases. This means the low-energy
cross section for $b = c$ does not maintain the universality. The only
way to keep the $R$-independence we should require the additional 
conditions
\begin{equation}
\label{addi}
W = - \frac{m^2 T^2}{n (a - n)}
\hspace{2.0cm}
a \geq 2 n.
\end{equation}

If $T$ is a real parameter, these additional conditions seem to change the 
Lorentz signiture in the near-horizon region and thus may generate a serious
causal problem. We guess this problem may be originated from the matching
between the solutions whose valid regions are too distant. If our guess is 
right, the problem may be cured by introducing the intermediate region
between near-horizon and asymptotic regions as Unruh did in 
Ref.\cite{unruh76}. This issue seems to need a careful treatment and we 
hope to discuss it in the future. In this paper we will not go further this
causal problem.

If one takes $\omega \rightarrow 0$ limit in 
Eq.(\ref{coeff4}) with making use of Eq.(\ref{addi}), the coefficients
$A$ and $B$ become
\begin{equation}
\label{coeff5}
A=0    \hspace{1.0cm}
B= \frac{(-1)^{\frac{n-1}{2}}}{2^{\frac{n}{2}} \omega^{\frac{a}{2} - n}}
\left( \frac{2 b}{\omega T} \right)^{\frac{a - 2 n}{2 b}}
\frac{i \Gamma \left(\frac{a - 2 n}{2 b} \right)}
     {\Gamma \left(\frac{n}{2}\right)}
    v^n.
\end{equation}
Thus the coefficients $A$ and $B$ are independent of the matching point
$R$ as expected.
Inserting Eq.(\ref{coeff5}) into Eq.(\ref{cross2}) makes the low-energy
cross section to be
\begin{equation}
\label{lcross2}
\sigma_{L}^{(2)} = \left( \frac{\omega T}{2 b} \right)^{\frac{a - 2n}{b}}
\frac{2^{2n+1} \pi^{\frac{n}{2}} n b S}{T \omega^{2n+1}}
\frac{\Gamma^3 \left( \frac{n}{2} \right)}
     {\Gamma^2 \left(\frac{a - 2n}{2 b} \right)}
     v^{-n}.
\end{equation}
It is interesting to note that $\sigma_{L}^{(2)}$ is proportional to $v^{-n}$
while $\sigma_{L}^{(1)}$ in Eq.(\ref{cross3}) is proportional to 
$v^n$. This inverse power makes $\sigma_{L}^{(2)}$ to exhibit an increasing
behavior with respect to mass $m$ unlike $\sigma_L^{(1)}$.  

If one takes a $\omega \rightarrow \infty$ limit in Eq.(\ref{coeff4}), one can
compute the high-energy cross-section $\sigma_H^{(2)}$ for $b = c$ case. 
Although $\omega \rightarrow \infty$ limit of the coefficients $A$ and 
$B$ are different from those in $b < c$, this difference is only phase 
factor in leading order and therefore does not change the high-energy
cross section, {\it i.e.} $\sigma_H^{(2)} = \sigma_H^{(1)}$. Thus the 
universality is not maintained in this case too, and the $\omega$-dependence
is $\sigma_H^{(2)} \propto \omega^{-(n+1)}$.

Now let us discuss the absorption cross section when $b = 2 c$. If one 
matches the asymptotic solution (\ref{solution2}) and the near-horizon
solution (\ref{solution5}) at low energy, 
one can show straightforwardly the coefficients
$A$ and $B$ become
\begin{equation}
\label{revise2}
A=i 2^{\frac{n}{2}} \Gamma\left(1 + \frac{n}{2}\right) D_{\ell} (\eta)
\left( \frac{\omega^{b + 1} T}{b} \right)^{-\frac{a - b}{2 b}}
                                           \hspace{1.0cm} 
B= 0
\end{equation}
where
\begin{equation}
\label{revise3}
D_{\ell} (\eta) = \frac{\Gamma(2 \ell + 1)}
                       {2^{\ell} e^{-\frac{\pi}{2} \eta}
                        |\Gamma(\ell + 1 + i \eta)|}.
\end{equation} 
Inserting (\ref{revise2}) and ${\cal F}_0^{in} = S / \omega^{a - b - 1}$, 
which can be computed after tedious calculation with making use of the 
asymptotic formula of the Coulomb wave function\cite{abra72}, into 
Eq.(\ref{cross1}), one can derive the following low-energy absorption 
cross section
\begin{equation}
\label{revise4}
\sigma_L^{(3)} = \left( \frac{\Gamma \left(\frac{a}{2 b} \right)}
                             {\sqrt{\pi} \omega^{a - b} 
                              \Gamma \left(1 + \frac{a}{b} \right)}
                                                            \right)^2
                 \left( \frac{4}{\omega^2 T^2} \right)^{\frac{a}{b} - 1}
                 e^{-\pi \eta}
                \bigg|\Gamma\left(\frac{a + b}{2 b} + i \eta \right)\bigg|^2
                \sigma_L^{(1)}
\end{equation}
where $\sigma_L^{(1)}$ is given in Eq.(\ref{cross3}). Using formulae
of the gamma function\cite{abra72},
the final form of $\sigma_L^{(3)}$ reduces to 
\begin{equation}
\label{revise6}
\sigma_L^{(3)} = \left( \frac{b}{a \omega^{a - b}} \right)^2
                 \left(\frac{1}{\omega^2 T^2} \right)^{\frac{a}{b} - 1}
                 e^{-\pi \eta} \sigma_L^{(1)}
                 \prod_{n=0}^{\infty}
                 \left[1 + \left(\frac{\eta}{\frac{a}{2 b} + n + \frac{1}{2}}
                                  \right)^2  \right]^{-1}.
\end{equation}
Roughly speaking, therefore, we can say that the low-energy absorption cross
section at $b = 2 c$ is a mutiplication of some damping factor to the 
low-energy cross section at $b < c$. This can be seen more clearly if $a = b$,
where $\sigma_L^{(3)}$ simply reduces to 
\begin{equation}
\label{revise7}
\sigma_L^{(3)} = \sigma_L^{(1)} \frac{(\pi \eta) e^{-\pi \eta}}{\sinh \pi \eta}
\end{equation}
where $\sigma_L^{(1)} = \Omega_{n+1} S v^n$. Thus if $\pi \eta << 1$ or 
$\pi \eta >> 1$, the damping factor becomes roughly $1 - \pi \eta$ or 
$2 (\pi \eta) e^{-2 \pi \eta}$ respectively.

The universal property of the low-energy absorption cross section is examined 
when the mass effect is involved. Taking an assumption (\ref{assump}) in the 
near-horizon region we have shown that the universal property is maintained
when $0 < b < c$. If, however, $b = c$, the universal property is in general 
broken unless Eq.(\ref{addi}) is imposed. At $b = 2 c$ the universal property
is maintained and the final expression of the absorption cross section is 
a multiplication of some damping factor to the cross section at $b < c$. 

We guess the pecular behavior at $b = c$ is originated from the consideration 
of S-wave approximation. To go beyond S-wave approximation we should involve
an angle-dependent term which generates an additional effective potential 
in the radial equation (\ref{schro}). This effective potential generally
enables us to make an intermediate region between asymptotic and near-horizon
regions, where the effective potential is dominant compared to other 
factors in the potential. 
Matching $\phi_{\omega}^0$ and $\phi_{\omega}^{\infty}$ {\it via}
the solution in the intermediate region may remove the pecular behavior at 
$b = c$. 

Another interesting work in this issue is to go beyond the low-energy
approximation. For the case of the $4d$ Schwarzschild black hole the absorption
cross section in the entire range of $\omega$ is computed in 
Ref.\cite{sanc77,sanc78} using a series solutions in the near-horizon
and the asymptotic regions\cite{persi73}. The most striking result when we 
study the absorption and emission problems of black hole from the viewpoint
of the scattering theories is that the partial scattering amplitude loses 
its unitary property, which is closely related to the information loss.
Thus one can apply the computational method of Ref.\cite{sanc77,sanc78} 
to the higher-dimensional theories to go beyond the low-energy approximation.
Work in this direction was done recently in Ref.\cite{kanti02,kanti03}. In this
way it may be possible to check our guess for the high-energy limit of 
the absorption cross section.


{\bf Acknowledgement}:  
This work was supported by the Kyungnam University
Research Fund, 2002.


\begin{figure}
\caption{The plot of potential $V_0$ in terms of the ``tortose'' coordinate.
Usually the potential $V_0$ makes a barrier which separates the asymptotic
and near-horizon regions.}
\end{figure}
\vspace{0.4cm}
\begin{figure}
\caption{Plot of $\sigma_L^{(1)}$-vs-$\omega$ with various matching points
when $m=0.01$, and $n=a=b=T=S=1$. This figure indicates the low-energy
absorption cross section is independent of the matching point, which is 
the origin of universality.} 
\end{figure}

\newpage
\epsfysize=10cm \epsfbox{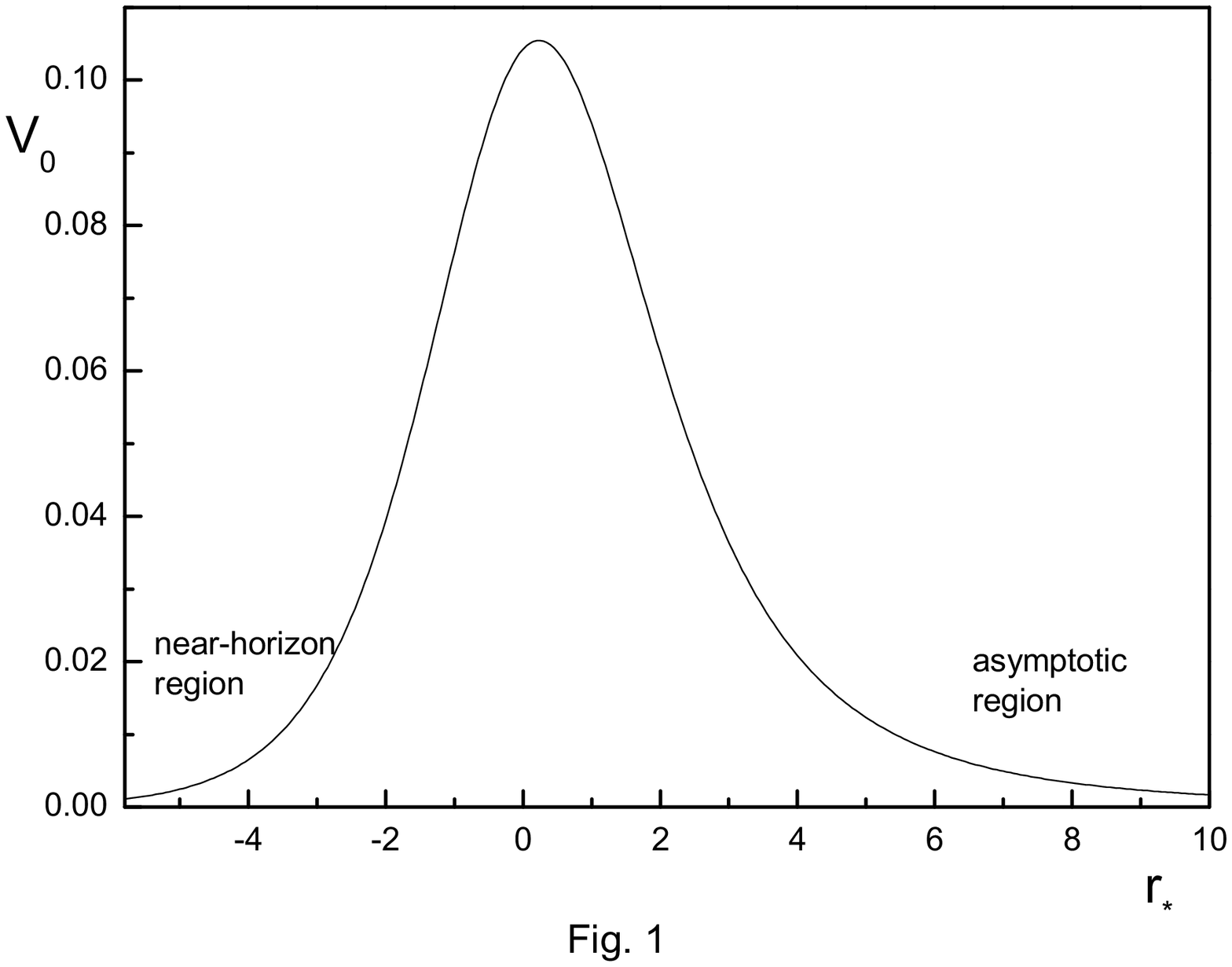}
\newpage
\epsfysize=10cm \epsfbox{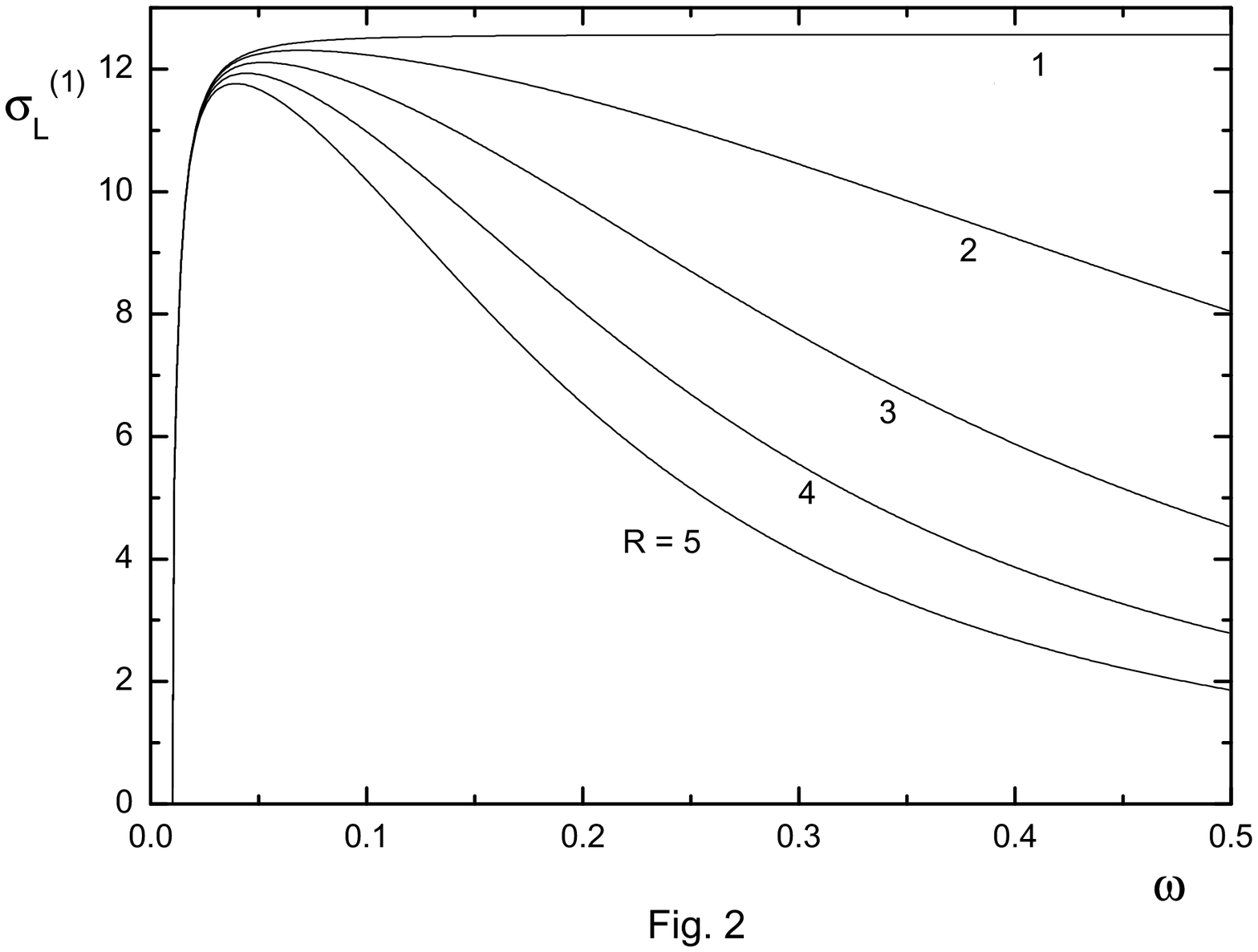}
\end{document}